\documentclass[aps,prl,groupedaddress,showpacs,notitlepage,twocolumn]{revtex4-1}
\usepackage{hyperref}
\usepackage{amsmath}
\usepackage{graphicx}
\usepackage{color}

\usepackage[T1]{fontenc}
\usepackage[latin1]{inputenc}

\newcommand{\be}{\begin{equation}}
\newcommand{\ee}{\end{equation}}

\newcommand{\bs}{\boldsymbol}
\newcommand{\br}{\bs{r}}
\newcommand{\bk}{\bs{k}}
\newcommand{\bn}{\bs{n}}

\begin{document}

\title{Anomalous Topological Phases and Unpaired Dirac Cones in Photonic Floquet Topological Insulators}

\author{Daniel Leykam$^1$}

\author{M. C. Rechtsman$^2$}

\author{Y. D. Chong$^{1,3}$}

\affiliation{$^1$Division of Physics and Applied Physics, School of Physical and Mathematical Sciences,\\
Nanyang Technological University, Singapore 637371, Singapore\\
$^2$Department of Physics, The Pennsylvania State University, University Park, PA 16802, USA\\
$^3$Centre for Disruptive Photonic Technologies, Nanyang Technological University, Singapore 637371, Singapore}

\date{\today}

\begin{abstract}
We propose a class of photonic Floquet topological insulators based on staggered helical lattices and an efficient numerical method for calculating their Floquet bandstructure. The lattices support anomalous Floquet topological insulator phases with vanishing Chern number and tunable topological transitions. At the critical point of the topological transition, the bandstructure hosts a single unpaired Dirac cone, which yields a variety of unusual transport effects: a discrete analogue of conical diffraction, weak antilocalization not limited by intervalley scattering, and suppression of Anderson localization. Unlike previous designs, the effective gauge field strength can be controlled via lattice parameters such as the inter-helix distance, significantly reducing radiative losses and enabling applications such as switchable topological wave-guiding.
\end{abstract}



\maketitle

Photonic topological insulators (PTIs) are an emerging class of photonic devices possessing topologically-nontrivial gapped photonic bandstructures \cite{topological_photonics_review,Raghu1,Raghu2,wang2009,khanikaev2013,rechtsman2013,hafezi2013,
liang2013,pasek2014,PQSH,microwave_network,plasmons,zeuner2015,PTCI}, analogous to single-particle electronic bandstructures of topological insulators \cite{topo_insulator_reviews}.  They have potential applications as robust unidirectional or polarization-filtered waveguides, and as scientific platforms for probing topological effects inaccessible in condensed-matter systems. In the technologically important optical frequency regime, only two PTIs have been demonstrated in experiment: arrays of helical optical waveguides \cite{rechtsman2013}, and coupled ring resonators \cite{hafezi2013,liang2013,pasek2014,microwave_network,plasmons}. These two different designs each possess unique advantages.  Waveguide array PTIs, for instance, allow the propagation dynamics of topological edge states to be directly imaged \cite{rechtsman2013}.  The design of the waveguide array PTI is based on the ``Floquet topological insulator'' concept~\cite{oka2009,lindner2011,gu2011,floquet_topological_insulators}, which originally described quantum systems with time-periodic Hamiltonians; the idea is that topologically nontrivial states can be induced via periodic driving \cite{oka2009,lindner2011,gu2011,floquet_topological_insulators}, rather than via magnetic or spin-orbit effects in a static Hamiltonian.  In the PTI, the Hamiltonian describes the classical evolution of the optical fields in the waveguide array, and its periodic drive arises from the helical twisting of the waveguides \cite{rechtsman2013,modulated_lattices}.

Floquet topological insulators are highly interesting because they exhibit topological phenomena that have no counterparts in static Hamiltonians~\cite{floquet_reviews,kitagawa2010,rudner2013,asboth2014,titum2015,carpentier2015,titum2015b,Wang2016}.  For example, there can exist two dimensional (2D) ``anomalous Floquet insulator'' (AFI) phases which are topologically nontrivial---including hosting protected edge states---despite all bands having zero Chern number \cite{kitagawa2010,rudner2013,asboth2014,titum2015,carpentier2015,microwave_network,plasmons}.  When disorder is introduced, the anomalous topological edge states become the only extended states, with all other states localized \cite{titum2015b}.  At critical points between topological phases, Floquet bandstructures can exhibit unpaired Dirac cones, defeating the ``fermion-doubling'' principle~\cite{dirac_network}.  It is thus noteworthy that these unusual features were \textit{not} accessed by the experiments of Ref.~\onlinecite{rechtsman2013}.  The waveguide array was always observed in a standard Chern insulator phase generated by weak periodic driving; transitions to any other topologically nontrivial phase were unachievable because the strength of the effective gauge field was controlled by the bending radius of the helical waveguides. Radiative losses, which increase exponentially with bending \cite{bending_loss}, came to dominate before any ``strong field'' topological transitions were reached~\cite{rechtsman2013}.

This paper describes a class of waveguide arrays overcoming the above limitations, allowing for the observation of topological transitions between conventional insulator, Chern insulator, and AFI phases, as well as unpaired Dirac cones at the transition points. To the best of our knowledge, AFI phases and unpaired Dirac cones have never been demonstrated in optical-frequency PTIs. Continuously tuned transitions between trivial and nontrivial topological phases, or into an AFI phase, have never been observed in any 2D PTI.  Our design is based on ``staggered'' lattices of helical waveguides, with each of the two sublattices having a different helix phase. The two-band Floquet bandstructure can be tuned to different topological phases by varying the nearest-neighbor coupling strength or sublattice asymmetry.  We can access different topological phases while maintaining small bending radii; the bending losses for reaching the conventional insulator to AFI transition are reduced by around two orders of magnitude compared to the topological transition discussed in Ref.~\onlinecite{rechtsman2013}.  This design thus shows promise for low-loss topological waveguides that are switchable (e.g.~via optical nonlinearity).

The critical Floquet bandstructure hosts an unpaired Dirac cone.  This is unlike any other previously observed photonic band-crossing points, which involve either paired Dirac points \cite{conical_diffraction}, quadratic dispersion \cite{chong2008}, or an attached flat band \cite{flatband,flatband2}. The unpaired Dirac cone is reminiscent of the chiral bandstructure of the Haldane model with broken parity and time-reversal symmetries~\cite{haldane1988}, or surface states of 3D topological insulators~\cite{topo_insulator_reviews,lu2015}.  Here, it arises from a Floquet process, and specifically the fact that the Floquet bandstructure is a ``quasienergy'' spectrum (see below).  Wave propagation in the critical PTI is immune to the inter-valley scattering that occurs with pairs of Dirac cones~\cite{ando2002}.  Based on this, we demonstrate a novel ``discrete'' conical diffraction effect, generated by exciting a single unit cell, as well as resistance to localization in the presence of short-range disorder \cite{bergmann1982}.  We note that although similar Floquet bandstructures that can host unpaired Dirac cones have previously been studied theoretically \cite{lindner2011,pasek2014,dirac_network}, those studies lacked information about the propagation dynamics of the Dirac cone states, which we can investigate using our experimentally realistic waveguide array models.

\begin{figure}
\includegraphics[width=\columnwidth]{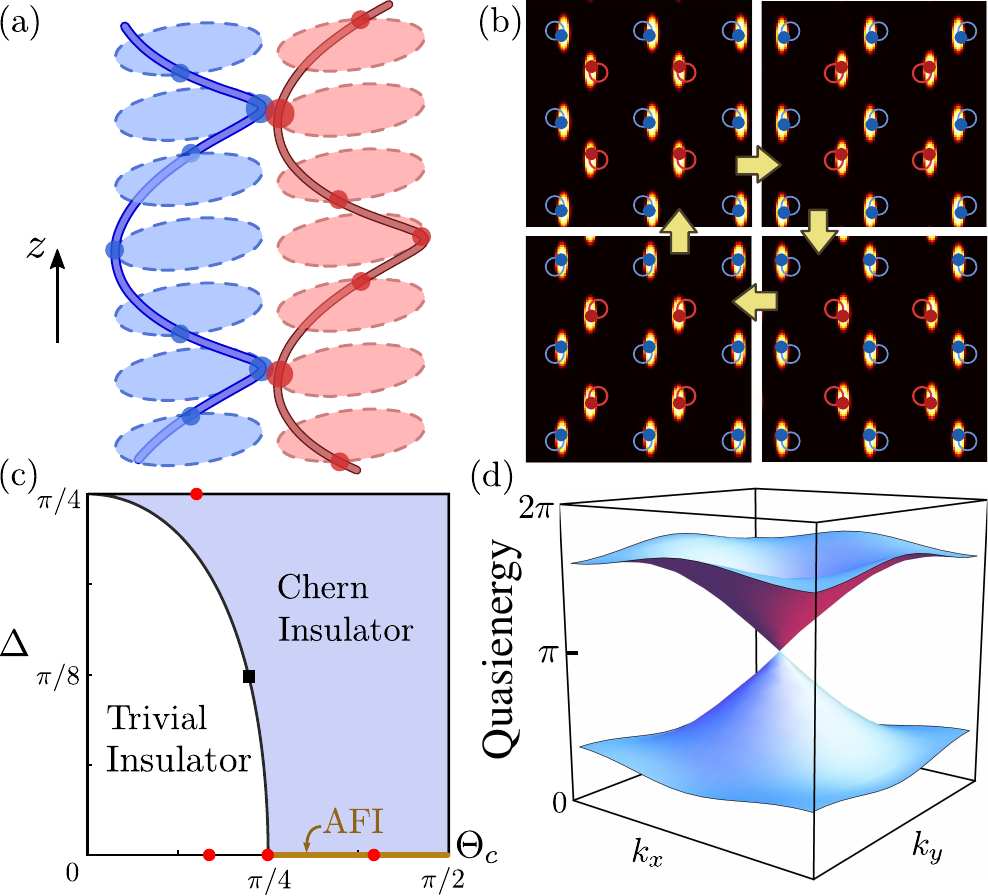}\\
\caption{(Color online) A square staggered lattice of helical waveguides. (a) Schematic of two neighboring waveguides, twisting clockwise along the propagation axis $z$ with relative phase shift $\pi$. (b) Cross section of the lattice potential at each helix quarter-cycle, with the circular trajectories overlaid. (c) Phase diagram of the tight-binding model, in terms of the sublattice asymmetry $\Delta$ and coupling strength $\theta_c$.  Red dots indicate the parameters for the band diagrams in Fig.~\ref{fig:band_structure}. The nontrivial phase forms an AFI along the $\Delta = 0$ line and a Chern insulator for $\Delta \ne 0$. (d) Tight-binding bulk spectrum at the black square in (c), along the phase boundary.}
\label{fig:lattice_diagram}
\end{figure}

An example of the staggered helix design is the square lattice shown in Fig.~\ref{fig:lattice_diagram}(a)--(b). There are two sublattices, forming a checkerboard pattern; the helices on each sublattice are shifted relative to each other in the $z$ direction, by half a helix cycle. This produces a $z$-dependent separation between waveguides, so that each waveguide approaches its four nearest neighbors in turn at each quarter-cycle.  Similar schemes can be implemented in other lattice geometries, such as a honeycomb lattice \cite{kitagawa2010}.  For simplicity, this paper focuses on the square lattice.


First, we model the lattice in a tight-binding approximation similar to a 2D discrete-time quantum walk~\cite{quantum_walks}.  Since the inter-waveguide couplings are evanescent, we assume each waveguide couples to one neighbor at a time. The Floquet evolution operator, $\hat{U}$, is defined by $\psi( z + Z) = \hat{U} \psi(z)$, where $Z$ is the helix period and $\psi = (\psi_A,\psi_B)$ are the tight-binding amplitudes on each sublattice. $\hat{U}$ factorizes into a series of independent two-waveguide couplings, separated by free evolution:
\begin{align}
  \hat{U} = \hat{S} \left( -k_- \right)  \hat{S} \left( -k_+ \right)  \hat{S} \left( k_- \right)  \hat{S} \left( k_+ \right), \label{eq:tight_binding}
\end{align}
with the notation $k_\pm \equiv (k_x \pm k_y)/\sqrt{2}$, where $k_{x,y}$ are the crystal momenta in units of the inverse waveguide separation in the absence of modulation; and
\begin{equation}
  \hat{S}(\kappa) = \left( \begin{array}{cc} e^{i \Delta} \cos \theta_c & -i e^{i (\Delta + \kappa)} \sin \theta_c  \\ -i e^{-i (\Delta + \kappa)} \sin \theta_c & e^{-i \Delta}\cos \theta_c \end{array} \right),
\end{equation}
where $\Delta$ is a small detuning between the sublattice propagation constants (which can be implemented by having different waveguide refractive indices), and $\theta_c$ is the coupling strength.  Since $\hat{U}$ is unitary, its eigenvalues have the form $e^{i \beta (\bk)}$ where $\beta(\bk)$ is the ``quasienergy'' spectrum.  Note that this model resembles the 2D quantum walk described in Ref.~\cite{rudner2013}, with time evolution replaced by propagation in $z$, and that $\hat{S}(\kappa)$ is the most general scattering matrix permitted by the lattice symmetries~\cite{supplementary}.

Fig.~\ref{fig:lattice_diagram}(c) shows the phase diagram of the quasienergy bandstructure, as a function of $\Delta$ and $\theta_c$.  The system is a trivial insulator at weak couplings, and a topological insulator above a critical coupling strength~\cite{supplementary}. At the transition, the bandstructure has an unpaired Dirac cone at the $\Gamma$ point, as shown in Fig.~\ref{fig:lattice_diagram}(d). Increasing $\Delta$ pushes the two bands away from quasienergy $\beta = 0$ and closer towards reconnecting at $\beta = \pm \pi$, reducing the critical coupling strength. At $\theta_c = \pi / 2$, the bands merge into a topological flat band~\cite{rudner2013}.

The $\Delta = 0$ case is particularly interesting.  Here, the sublattice symmetry enforces a line degeneracy at the Brillouin zone edge, so there is a single band gap.  For small $\theta_c$, the spectrum resembles that of an unmodulated square lattice with a single Bloch band folded back onto itself. At the critical point $\theta_c = \pi / 4$, the formation of the Dirac cone leads to a completely gapless spectrum.  A long-wavelength expansion of $\hat{U} \approx \exp[ -i (\hat{H}_D - \pi)]$ about the $\Gamma$ point yields an effective Dirac Hamiltonian,
\be 
\hat{H}_{D} (\bk ) = -k_x \hat{\sigma}_z + k_y \hat{\sigma}_y - 4 (\theta_c - \pi / 4) \hat{\sigma}_x, \label{eq:dirac}
\ee
where $\hat{\sigma}_{x,y,z}$ are the Pauli matrices. For $\theta_c > \pi/4$, the system is an AFI \cite{kitagawa2010,rudner2013,asboth2014,titum2015,carpentier2015} with unidirectional topological edge states.

To apply these ideas to a realistic photonic lattice, such as femtosecond laser-written waveguides in fused silica \cite{rechtsman2013,titum2015,femtosecond_arrays}, we now go beyond the tight-binding description.  A photonic lattice is described by a paraxial field $\psi(\br, z)$ governed by the Schr\"odinger equation
\be
i \partial_z \psi = -\frac{1}{2k_0} \nabla_{\perp}^2 \psi - \frac{k_0 \delta n (x,y,z)}{n_0} \psi, \label{eq:schrodinger}
\ee
where $\nabla_{\perp}^2 = \partial_x^2 + \partial_y^2$, $k_0 = 2 \pi n_0 / \lambda$, and the refractive index is $n_0 = 1.45$ at wavelength $\lambda = 633$nm, with modulation $\delta n \sim 7.5\times 10^{-4}$. Similar to real experiments, we give the waveguides elliptical cross sections with axis diameters $11\mu$m and $4\mu$m~\cite{rechtsman2013}, as shown in Fig.~\ref{fig:lattice_diagram}(b).  They form a square lattice with mean waveguide separation $a$, helix radius $R_0$, and pitch $Z$. We can increase the effective coupling, $\theta_c$, by increasing $1/a$, $R_0$, or $Z$. 

Direct calculation of the Floquet bandstructure for a continuum model (as opposed to a tight-binding model) is a nontrivial task, because the quasienergies $\beta_{n,\bk}$ are defined modulo $2\pi/Z$, so there is no ground state for numerical eigensolvers to converge on, and continuum (unguided) modes enter in an uncontrolled way. We devised an efficient method for performing this calculation by truncating the evolution operator $\hat{U}$ to a basis formed by the \emph{static} Bloch waves at $z=0$. This amounts to a quasi-static approximation neglecting coupling to unbound (continuum) modes.  Bending losses can be estimated via the norm of the Floquet evolution operator eigenvalues, by findings its deviation from unitarity.  Further details are given in the Supplemental Material \cite{supplementary}.

We now fix $R_0 = 3\,\mu$m and $Z=2$\,cm.  This yields a loss of $\lesssim 0.02$dB/cm, independent of topological phase, which we tune by varying $a$ and/or $\Delta$.  By constrast, the strength of the effective gauge field in the unstaggered lattice of Ref.~\onlinecite{rechtsman2013} was tuned by increasing $R_0$, which also increased the bending losses exponentially~\cite{bending_loss}.  That limited the system to the ``weak field'' perturbative regime; losses exceeded 3dB/cm before reaching a predicted strong field topological transition (between two Chern insulators), making that transition unobservable.

\begin{figure}
\includegraphics[width=\columnwidth]{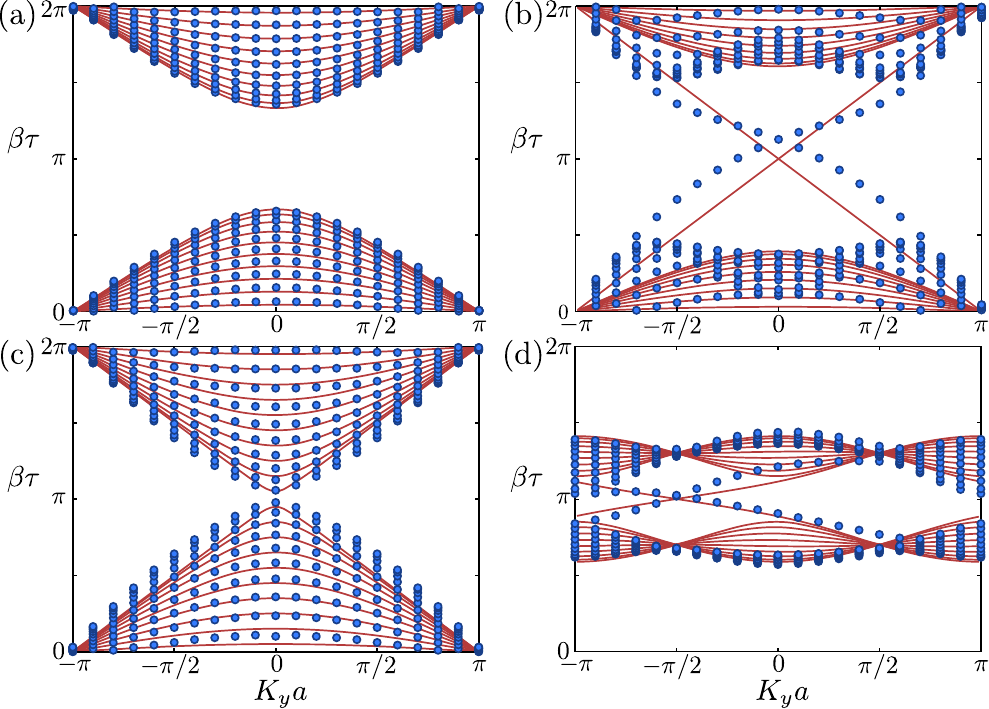}
\caption{(Color online) Band structures for a semi-infinite strip $10$ unit cells wide. Blue points are obtained from the continuum model, and red curves from the tight-binding model. (a) Trivial insulator ($a=25\mu$m, $\theta_c \approx 0.17\pi, \Delta = 0$). (b) Anomalous Floquet insulator ($a=20\mu$m, $\theta_c \approx 0.4 \pi, \Delta = 0$). (c) Critical phase ($a = 23\mu$m, $\theta_c \approx \pi/4, \Delta = 0$). (d) Chern insulator ($a=23\mu$m, $\theta_c \approx 0.15\pi, \Delta \approx \pi/4$).}
\label{fig:band_structure}
\end{figure}


Fig.~\ref{fig:band_structure} shows the band structure for a strip geometry. For comparison, results from the truncated-Bloch method are plotted together with the results from a fitted tight-binding model~\cite{supplementary}. The two methods agree well, particularly in the weak-coupling regime. In Fig.~\ref{fig:band_structure}(a), we see that the system is a trivial insulator, with a single band and a single gap (note that the spectrum is periodic along the $\beta$ axis), whereas in Fig.~\ref{fig:band_structure}(b), the gap has closed and reopened, inducing chiral edge states centered at $\beta \sim \pi$. This is the AFI phase; the Chern number of the single band is necessarily zero, despite the presence of chiral edge states. The transition point is shown in Fig.~\ref{fig:band_structure}(c), which features an unpaired Dirac point at the center of the Brillouin zone.  Fig.~\ref{fig:band_structure}(d) shows a $\Delta\neq 0$ case, corresponding to a Chern insulator; there is both a trivial gap and a nontrivial gap, and the two bands have Chern numbers 1 and -1, as in the Haldane model~\cite{chern_numerical}.

\begin{figure}
\includegraphics[width=\columnwidth]{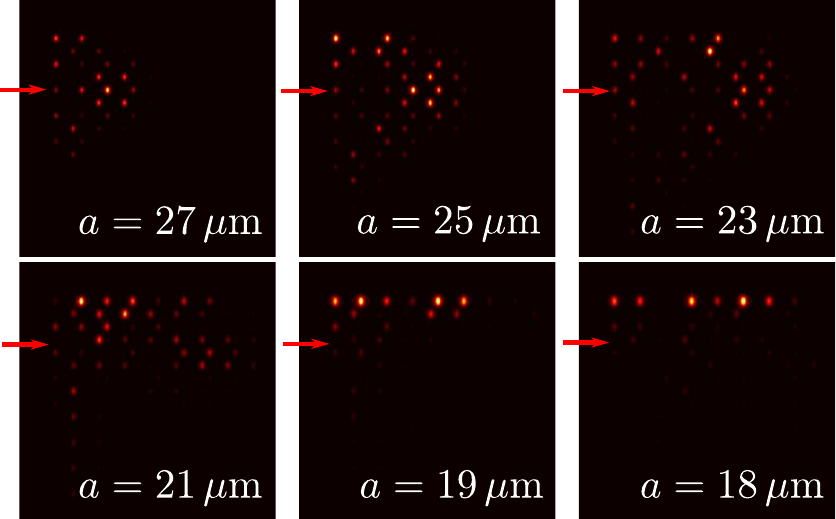}
\caption{(Color online) Output intensity profile after progation through $5Z$, with one edge site initially excited (red arrow).  Reducing the lattice period causes a transition into an anomalous Floquet insulator phase with topological edge states.}
\label{fig:edge_propagation}
\end{figure}

The topological transitions can be probed via beam propagation experiments. Fig.~\ref{fig:edge_propagation} shows beam propagation simulations with a single waveguide initially excited along the edge. For large $a$, with the lattice in the trivial phase, the excitation simply spreads into the bulk. Upon decreasing $a$, we observe a strongly-localized mode that propagates unidirectionally along the edge, including around corners.  This is a clear signature of a topological transition to the AFI, which has never been experimentally demonstrated in a 2D photonic lattice. We stress that varying $a$ is just one of many possible tuning methods. Due to the strong sensitivity of the evanescent coupling strength to waveguide mode localization, there are other interesting ways to achieve controllable switching between topological phases, such as the Kerr effect or thermal tuning.

\begin{figure}

\includegraphics[width=\columnwidth]{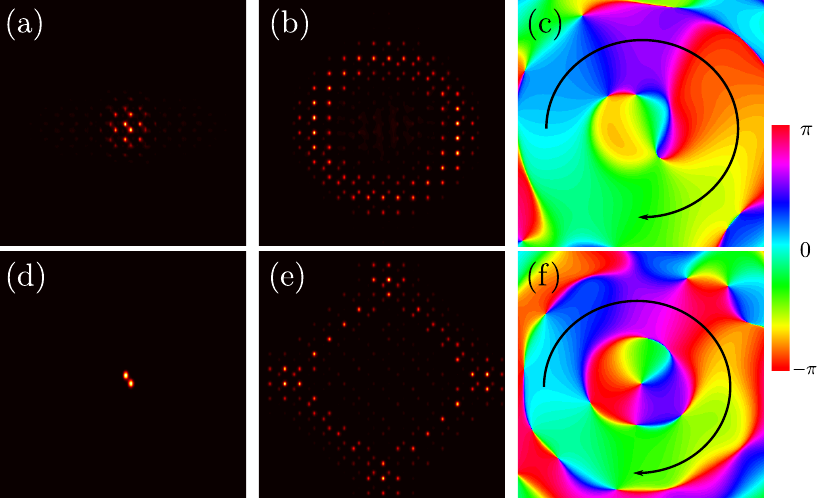}

\caption{(Color online) Conical diffraction arising from the unpaired Dirac point.  In the upper panels, the initial excitation is Gaussian; in the lower panels, only one unit cell is excited.  (a,d) Input and (b,e) output beam intensity. (c,f) Phase profile of output cross-polarized pseudospin component, showing clockwise phase circulation (black arrows). The lattice size is $16\times16$ unit cells, and the propagation distance is $L = 6Z$.}

\label{fig:conical_intensities}
\end{figure}

It is also interesting to study the behavior of the lattice at the critical point of the topological transition, where the quasienergy bandstructure contains an unpaired Dirac cone at the Brillouin zone center. A direct method for revealing the existence of a Dirac cone is ``conical diffraction'', which involves constructing an initial wavepacket from Dirac cone states, which then evolves (under linear relativistic dispersion) into a ring with constant thickness and nonzero phase winding.  In honeycomb lattices with two Dirac cones, conical diffraction requires selectively exciting one cone, e.g.~using a tilted spatially-structured input beam \cite{conical_diffraction}. With an unpaired Dirac cone, however, we can generate conical diffraction using simple \emph{unstructured} Gaussian beams at normal incidence, as shown in Fig.~\ref{fig:conical_intensities}(a)--(b).  This exclusively excites ``pseudospin-up'' Dirac modes governed by Eq.~\eqref{eq:dirac}, with chirality determined by the chirality of the modulation $\delta n(\br,z)$.  This intrinsic chirality is revealed by the phase of the diffracted field. Pseudospin angular momentum generates an optical vortex in the ``cross-polarized'' pseudospin-down component of the diffracted field, with vortex charge sensitive to the chirality of the Dirac dispersion~\cite{vortex_generation}. Here, ``pseudospin-down'' corresponds to light scattered into the second Brillouin zone, readily measured via Fourier filtering~\cite{supplementary}. Fig.~\ref{fig:conical_intensities}(c) shows the phase profile, exhibiting the predicted topological charge.


What happens as we reduce the width of the initial Gaussian excitation? One might expect conical diffraction to be destroyed, since Eq.~\eqref{eq:dirac} is based on an effective-mass (carrier-envelope) approximation in the transverse plane. While that is the case for static Hamiltonians, here diffraction is preserved by the unique features of the Floquet bandstructure: the spectrum is entirely gapless, and has no local band maxima or minima. Consequently, the band velocity is nonzero almost everywhere, and the initial excitation evolves into a \emph{discrete} conical-like diffraction pattern with a dark central spot and nonzero vortex charge.  As shown in Fig.~\ref{fig:conical_intensities}(d)--(f), this holds true even when the initial excitation is reduced to a single unit cell.

\begin{figure}

\includegraphics[width=\columnwidth]{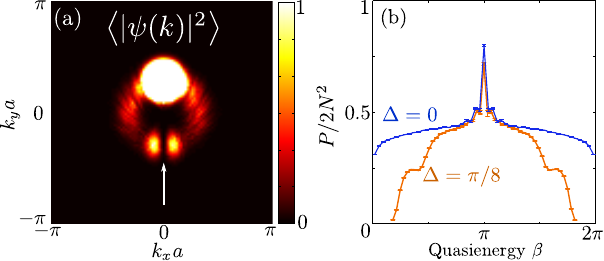}

\caption{(Color online) Disorder-insensitivity at the critical point. (a) Fourier intensity of a broad (width $5a$) probe beam, after propagating $60Z$ through a weakly-disordered \cite{supplementary} $100\times100$ lattice, averaging over 20 disorder realizations.  A weak antilocalization dip occurs in the backscattering direction (white arrow); the color scale saturates in the forward direction. (b) Participation number $P$ normalized by number of waveguides $2N^2 = 1800$, for the tight-binding eigenmodes of a strongly-disordered lattice.  Localization is absent for $\Delta = 0$, but Lifshitz localization tails appear for $\Delta = \pi/8$. }

\label{fig:disorder}

\end{figure}

Wave propagation at the critical point should be intrinsically robust against disorder, due to the enforced chirality and absence of band edges.  To show this, we introduce random site-to-site fluctuations in the waveguide detunings of the tight-binding model \eqref{eq:tight_binding}. For weak disorder, Dirac modes experience suppressed backscattering, a phenomenon known as ``weak antilocalization'' \cite{bergmann1982}.  Usually, weak antilocalization disappears when the disorder is short-ranged, due to inter-valley scattering~\cite{ando2002}.  However, Fig.~\ref{fig:disorder}(a) shows that weak antilocalization persists in our system even for completely short-range (site-specific) disorder. Furthermore, Anderson localization normally sets in at large disorder strengths, commencing at the band edges.  In Fig.~\ref{fig:disorder}(b), we probe the localization of the tight-binding eigenmodes by their mode participation numbers, and find that localization is defeated in the critical $\Delta = 0$ system due to the lack of band edges.  For $\Delta \ne 0$, the Floquet bandstructures have well-defined band edges, and we correspondingly observe Lifshitz tails of strongly-localized modes~\cite{supplementary}.

In summary, we have shown how to realize Floquet PTIs in staggered helical waveguide arrays. Novel topological transitions, beyond those characterized by Chern numbers, can be accessed by tuning lattice parameters other than the bending radius; this allows for low-loss operation and raises the prospect of nonlinear or actively-controllable, robust topological waveguide devices~\cite{TI_solitons}. The many interesting behaviors of the unpaired Dirac cone at the critical point, including discrete conical diffraction and suppression of Anderson localization, are worth probing in detail in future experiments.

\begin{acknowledgments}
We are grateful to N.~H. Lindner and H.~Wang for helpful discussions. This research was supported by the Singapore National Research Foundation under grant No.~NRFF2012-02, and by the Singapore MOE Academic Research Fund Tier 3 grant MOE2011-T3-1-005.  M.C.R. acknowledges the support of the National Science Foundation under grant No.~ECCS-1509546.
\end{acknowledgments}

\clearpage

\begin{center}
  {\large \textbf{Supplemental Material}}

  for

  {\large Anomalous Topological Phases and Unpaired Dirac Cones in Photonic Floquet Topological Insulators}

  {\footnotesize Daniel Leykam, M.~C. Rechtsman, and Y.~D.~Chong}
\end{center}

\makeatletter 
\renewcommand{\theequation}{S\arabic{equation}}
\makeatother
\setcounter{equation}{0}

\makeatletter 
\renewcommand{\thefigure}{S\@arabic\c@figure}
\makeatother
\setcounter{figure}{0}

\section{Tight-binding model}
\label{sec:tight_binding}

Here, we present a detailed derivation of the tight-binding model of the staggered square lattice PTI, leading to Eqs.~(1)--(2) of the main text.  Consider a generic two-waveguide coupler, which is described by the coupled mode Hamiltonian~\cite{supp_modulated_lattices},
\be 
\hat{H}_{\mathrm{coupler}} = \left( \begin{array}{cc} \omega & C \\ C & -\omega \end{array} \right),
\ee
where $C$ is the coupling strength and $\omega$ is the mode detuning. Reciprocity requires that the coupling $C$ be purely real (one can see this by rotating the coupler by $\pi$, which reverses the apparent waveguide velocities while representing the same physical system, so $C = C^*$). Hence the relative velocity between the two helical waveguides can at most renormalize the magnitude of $C$. Note that this reasoning only applies to nearest neighbor interactions. Next nearest neighbor coupling (between waveguides belonging to the same sublattice) can be analyzed using the frame transformation method of Ref.~\cite{supp_rechtsman2013} and displays a time-dependent complex phase.

The evolution operator for propagation by a distance $L$ is, to first order in $\omega$,
\begin{multline}
\hat{U}_{\mathrm{coupler}} = e^{-i \hat{H}_{\mathrm{coupler}} L}  \\
= \left( \begin{array}{cc} \cos (LC) - \frac{i\omega}{C} \sin (LC) & -i \sin (LC) \\ -i \sin (LC) & \cos (LC) + \frac{i\omega}{C} \sin (LC) \end{array} \right).
\end{multline}
As a simplifying assumption (which does not affect the topological properties of the model~\cite{supp_rudner2013}), we consider a vanishing interaction length $L \rightarrow 0$ with $C \rightarrow \infty$ such that the coupling strength is encoded by the dimensionless parameter $LC \rightarrow \mathrm{constant} = \theta_c$, i.e.~the coupler's principal Euler angle. Then the detuning $(\omega / C)\approx 0$ has a negligible effect during the interaction length.  A phase difference between the waveguides $\Delta \approx \omega Z / 4$ only accumulates during the ``free evolution'' parts of the cycle, when the waveguides are decoupled. Under these approximations, the scattering matrix describing the evolution of the field over one quarter cycle $Z/4$ is
\begin{align}
  \begin{aligned}
\hat{S}_0 &= \left( \begin{array}{cc} e^{i \Delta} & 0 \\ 0 & e^{-i\Delta} \end{array} \right) \hat{U}_{\mathrm{coupler}} \\
&=  \left( \begin{array}{cc} e^{i \Delta} \cos \theta_c & -i e^{i \Delta} \sin \theta_c  \\ -i e^{-i \Delta} \sin \theta_c & e^{-i \Delta}\cos \theta_c \end{array} \right). \label{eq:s0}
  \end{aligned}
\end{align}
We can now extend this treatment to the full 2D lattice.  Each unit cell consists of two sublattices, with wave amplitudes $a_{\bn}$ and $b_{\bn}$. Here $\bn = (n,m)$ indexes the unit cells, and we consider a lattice in the ``diamond'' configuration illustrated in Fig.~1 of the main text.  The two sublattices are displaced by $\bs{a} = \frac{a}{\sqrt{2}}(1,1)$ and the lattice vectors are $\bs{d}_1 = a \sqrt{2} (1,0)$, $\bs{d}_2 = a \sqrt{2} (0,1)$.

To derive the Bloch wave spectrum, we take the \textit{ansatz} $(a_{\bn},b_{\bn}) = (a_0,b_0 e^{i \bk \cdot \bs{a}} ) e^{i \bk \cdot \bn a \sqrt{2}}$, with crystal momenta $k_{x,y} \in (-\frac{\pi}{a\sqrt{2}},\frac{\pi}{a\sqrt{2}})$. A complete clockwise cycle consists of the sequence
\begin{subequations}
\begin{align}
\left( \begin{array}{c} a_1 \\ b_1 e^{i \bk \cdot \bs{a} } \end{array} \right) &= \hat{S}_0 \left( \begin{array}{c} a_0 \\ b_0 e^{i \bk \cdot \bs{a} } \end{array} \right), \\
\left( \begin{array}{c} a_2 \\ b_2 e^{i \bk \cdot (\bs{a}-\bs{d}_1) } \end{array} \right) &= \hat{S}_0 \left( \begin{array}{c} a_1 \\ b_1 e^{i \bk \cdot (\bs{a}-\bs{d}_2) } \end{array} \right), \\
\left( \begin{array}{c} a_3 \\ b_3 e^{i \bk \cdot (\bs{a}-\bs{d}_2) } \end{array} \right) &= \hat{S}_0 \left( \begin{array}{c} a_2 \\ b_2 e^{i \bk \cdot (\bs{a}-\bs{d}_1) } \end{array} \right), \\
\left( \begin{array}{c} a_0 \\ b_0 e^{i \bk \cdot (\bs{a}+\bs{d}_1) } \end{array} \right) &= \hat{S}_0 \left( \begin{array}{c} a_3 \\ b_3 e^{i \bk \cdot (\bs{a}+\bs{d}_2) } \end{array} \right),
\end{align}
\end{subequations}
which describes the sequential coupling between a waveguide and its four neighbors. It is convenient to combine the Bloch phase factors $\exp(i \bk \cdot \bs{a}) = \exp(i \kappa)$ with $\hat{S}_0$ into the scattering matrix,
\be 
\hat{S} (\kappa) = \left( \begin{array}{cc} e^{i \Delta} \cos \theta_c & -i e^{i (\Delta + \kappa)} \sin \theta_c  \\ -i e^{-i (\Delta + \kappa)} \sin \theta_c & e^{-i \Delta}\cos \theta_c \end{array} \right),
\ee
where $\kappa$ is the relative phase. $S(\kappa + \pi) = \hat{\sigma}_z S(\kappa) \hat{\sigma}_z$ is periodic ($\hat{\sigma}_{x,y,z}$ are the Pauli matrices). The evolution operator $\hat{U}(\bk)$ can be compactly written as
\begin{equation}
\hat{U}(\bk ) = \hat{S} \left(-k_-\right)  \hat{S} \left(-k_+\right)
\hat{S} \left(k_-\right)  \hat{S}\left(k_+\right),
\label{Floquet_sup}
\end{equation}
where we have defined
\begin{equation}
  k_\pm = a (k_x \pm k_y) /\sqrt{2}.
\end{equation}
From Eq.~(\ref{Floquet_sup}), we can find the quasienergy spectrum by
\be 
\hat{U}(\bk ) \begin{pmatrix}a_0(\bk) \\ b_0(\bk) \end{pmatrix}
= e^{-i \beta(\bk)} \begin{pmatrix}a_0(\bk) \\b_0(\bk) 
\end{pmatrix},
\ee
where $\beta(\bk)$ is the quasienergy.

\subsection{Symmetric lattice ($\Delta = 0$)}

When $\Delta = 0$, the lattice is invariant under the interchange of the two sublattices (described by the $\hat{\sigma}_x$ operation). The two bands are degenerate along the Brillouin zone edge, and for small $\theta_c$ the band structure $\beta \approx \pm 2 \theta_c (\cos k_+ +  \cos k_-)$ is equivalent to the dispersion of an ordinary square lattice with its single band folded back onto itself. Intuitively this is because in the weak coupling limit, the difference between sequential and simultaneous coupling becomes negligible, so the staggering of the helices is irrelevant.

\begin{figure}

\includegraphics[width=\columnwidth]{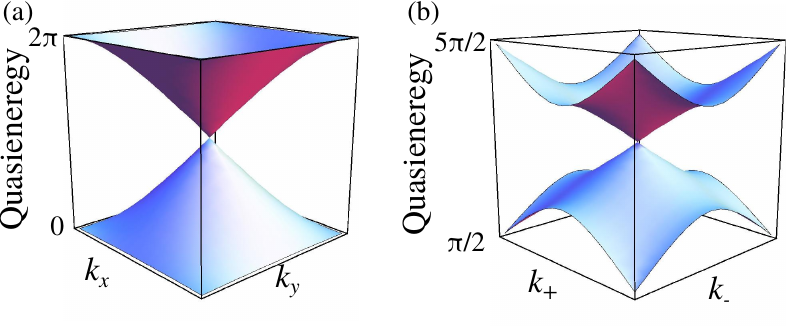}

\caption{(a) Gapless Floquet spectrum at $\theta_c = \pi/4$ of $\hat{U}(\bk)$ and (b) unfolded spectrum of $\hat{U}_{1/2}(k_{\pm})$, revealing the same structure as the model of Refs.~\cite{supp_liang2013,supp_dirac_network} with an additional Dirac cone appearing at the corner of the extended Brillouin zone $k_{\pm} \in [-\pi,\pi]$.} 
\label{fig:gapless_spectra}

\end{figure}

As $\theta_c$ is increased to $\pi / 4$, band edges at the $\Gamma$ point approach $\beta = \pm \pi$, reconnecting to form the ``critical'' gapless band structure shown in Fig.~\ref{fig:gapless_spectra}(a). To reveal the long wavelength dynamics near this critical point, we expand $\hat{U}$ about the $\Gamma$ point to first order in $\bk$ and $\delta \theta = \theta_c - \pi/4$,
\be 
\hat{U}_{\Gamma} \approx \exp \left\{ -i [\hat{H}_{D} (a \bk ) - \pi] + ... \right\},
\ee
where
\be 
\hat{H}_{D} (\bk ) = -k_x \hat{\sigma}_z + k_y \hat{\sigma}_y - 4 \delta \theta \hat{\sigma}_x, 
\ee
is an effective Dirac Hamiltonian with mass $4\delta \theta$, group velocity $v_F = 1$, and associated ``pseudospin'' $\hat{\sigma}_x$. The eigenstates of $\hat{\sigma}_x$, which are $(a,b) = (1,\pm 1)/\sqrt{2}$, involve excitations of both sublattices with equal intensity, in contrast to the pseudospin eigenstates in graphene which excite only a single sublattice. In the eigenbasis of $\hat{\sigma}_x$ and using polar coordinates $(k_x,k_y) = (k\cos \varphi,k\sin\varphi)$, the Dirac Hamiltonian takes the canonical form
\be 
\hat{H}_D(\bk ) = \left( \begin{array}{cc} - 4 \delta \theta & k e^{- i \varphi} \\ k e^{i \varphi} & 4 \delta \theta \end{array} \right). \label{eq:dirac_hamiltonian}
\ee
This effective Dirac Hamiltonian has the chiral symmetry $\hat{\Gamma} = 2 \hat{S}_z$. The evolution operator $\hat{U}$ shares this symmetry~\cite{supp_asboth2014},
\be 
\hat{\Gamma} \hat{U} \hat{\Gamma} = \hat{U}^{-1},
\ee
to first order in $\bk$. Thus, even though the full system does not have any obvious chiral symmetry, the symmetry is locally restored at the Dirac point.

Both the full band structure and effective Dirac Hamiltonian of the symmetric lattice resemble the ring resonator model of Refs.~\cite{supp_liang2013,supp_dirac_network}. In fact, there is an exact correspondence between the two: When $\Delta = 0$, the scattering matrix $\hat{S}(\kappa)$ has the symmetry $\hat{\sigma}_x \hat{S}(\kappa) \hat{\sigma}_x = \hat{S}(-\kappa)$, and $\hat{U}$ factorizes into
\begin{align}
\hat{U} &= \hat{\sigma}_x \hat{S}(k_-) \hat{\sigma}_x \hat{\sigma}_x \hat{S}(k_+) \hat{\sigma}_x \hat{S}(k_-) \hat{S}(k_+) \nonumber \\
&= \left[\hat{\sigma_x} \hat{S}(k_-) \hat{S}(k_+) \right]^2\nonumber \\
&\equiv \hat{U}_{1/2}^2.
\end{align}
If one considers the unfolded Brillouin zone $k_{\pm} \in [-\pi,\pi]$, the spectrum of $\hat{U}_{1/2}$ shown in Fig.~\ref{fig:gapless_spectra} is the same (up to a global phase shift) as the Floquet operators appearing in the network model of Ref.~\cite{supp_dirac_network}. The main difference between the two is that in our helix lattice the degeneracy between the two sublattices is lifted at each point in time, and it is only restored when one considers the evolution operator for a complete cycle. This explains the line degeneracy in Fig.~\ref{fig:gapless_spectra}(a): the eigenvalues ``see'' a square lattice with period $a$, while the symmetry-breaking instantaneous potential increases the actual lattice period to $a\sqrt{2}$, folding the Brillouin zone back onto itself. This zone folding is what allows the ``single'' Bloch band of the square lattice to reconnect with itself to form the anomalous Floquet topological insulator. 

\subsection{Asymmetric lattice ($\Delta \ne 0$)}

For $\Delta \ne 0$, the sublattice symmetry is broken so $\hat{U}$ no longer factorizes into $\hat{U}_{1/2}$ and there is no longer any direct correspondence with the network model of Refs.~\cite{supp_liang2013,supp_dirac_network}. The broken symmetry lifts the line degeneracy at the Brillouin zone edge, opening a topologically trivial gap. Because the two bands are pushed apart, the formation of the $\Gamma$ point Dirac cone is shifted to the weaker coupling strength (cf. Fig.~1 in main text),
\be 
\theta_c = \cos^{-1} [ \sec (\Delta )/\sqrt{2} ] \le \pi / 4. \label{eq:critical_coupling}
\ee
This critical coupling now marks the transition to a Chern insulator phase because there is a second, trivial band gap. Nevertheless the long wavelength dynamics at this critical point remain almost the same as the $\Delta = 0$ case: performing a similar series expansion to the one above, we obtain an effective Dirac Hamiltonian with a $\Delta$-dependent group velocity $v_F = \sec (\Delta) \sqrt{ \cos (2 \Delta )}$ and a rotated pseudospin,
\be 
\hat{S}_z(\Delta) = \frac{1}{2} (\hat{\sigma}_x \cos \Delta - \hat{\sigma}_y \sin \Delta ) \sqrt{1 - \tan^2 \Delta} -\frac{1}{2} \hat{\sigma}_z \tan \Delta,
\ee
continuously interpolating between $\hat{S}_z(0) = \frac{1}{2}\hat{\sigma}_x$ and $\hat{S}_z(\frac{\pi}{4}) = -\frac{1}{2}\hat{\sigma}_z$, resembling the pseudospin in graphene. This pseudospin is derived by introducing the total angular momentum operator $\hat{J}_z = \hat{L}_z + \hat{S}_z$, where $\hat{L}_z = i \partial_{\varphi}$ is the usual orbital angular momentum operator, and enforcing its conservation via $[\hat{J}_z,\hat{U}]=0$, which has a nontrivial solution because the spectrum has rotational symmetry.

Increasing $\theta_c$ further, the bands become increasingly flat before the spectrum becomes completely degenerate at $\theta_c = \pi/2$, corresponding to $\hat{U} = \hat{1}$. Similar to Ref.~\cite{supp_rudner2013}, this second gap closing marks the formation of an ``anomalous'' Floquet insulator phase, with edge modes traversing the remaining band gap. Since the bulk evolution operator is trivial (simply the identity), new topological invariants specific to Floquet systems are required to describe this phase.

\section{Continuum model}
\label{sec:continuum_model}

Our continuum model uses parameters similar to Refs.~\cite{supp_rechtsman2013,supp_titum2015}, describing femtosecond laser-written waveguide arrays in fused silica~\cite{supp_femtosecond_arrays}. Propagation is governed by the paraxial (Schr\"odinger) equation,
\be 
i \partial_z \psi = -\frac{1}{2k_0} \nabla^2 \psi - \frac{k_0 \Delta n (x,y,z)}{n_0} \psi,
\ee
where $k_0 = 2 \pi n_0 / \lambda$, the background refractive index is $n_0 = 1.45$ at wavelength $\lambda = 633$nm, and the index modulation forms a square helix lattice:
\begin{multline}
\Delta n(x,y,z) = \Delta n_1 \sum_{nm} \Big\{ V_0(X_{n}^-, Y_{m}^-) \\
+ \alpha V_0 (X_{n+1/2}^+, Y^+_{m+1/2}) \Big\}, \label{eq:lattice_potential}
\end{multline}
where
\begin{align}
X_{n}^\pm(z)  &= x \pm x_0(z) - nd \\
Y_{m}^\pm(z) &=y \pm y_0(z) - md.
\end{align}
$\Delta n_1 = 7.5 \times 10^{-4}$ is the modulation depth, $\alpha$ is the asymmetry between the sublattice depths, and $d=a\sqrt{2}$ is the lattice period (recall that $a$ is the separation between the sublattices). Putting lattice in the ``diamond'' configuration minimizes the effect of the $x-y$ anisotropy of the individual waveguide profiles $V_0$, described by the hypergaussian function
\be 
V_0(x,y) = \exp \left( - [(x/\sigma_x)^2 + (y / \sigma_y )^2 ]^3 \right),
\ee
with widths $\sigma_x = 2\,\mu$m and $\sigma_y = 5.5\,\mu$m. Using these parameters, representative values of the coupling constant between neighboring waveguides range from $C=2.7/$cm (for separation $12\mu$m) to $C=0.1/$cm (for separation $27\mu$m).

The waveguide centers move in helices according to
\be 
x_0(z) = R_0  \cos (\Omega z ), \quad y_0(z) =  R_0 \sin (\Omega z),
\ee
where $R_0 = 3 \mu$m is the helix radius, and $\Omega = 2 \pi / Z$ with pitch $Z = 2$cm. The helices of the two sublattices are staggered and $\pi$ out of phase, as given by Eq.~\ref{eq:lattice_potential}. The minimum feasible center-to-center waveguide separation is approximately $12\,\mu$m~\cite{supp_titum2015} (otherwise the waveguides would overlap and coalesce), leading to the constraint $a \ge 12\mu\mathrm{m} + 2R_0 = 18\mu$m. The helix radius and pitch are chosen to minimize the waveguide acceleration and associated bending losses. Array lengths of $L = 10-15$cm are feasible, corresponding to $5-7$ complete modulation cycles used in the main text.  The number of cycles can be further increased by reducing the pitch, at the expense of stronger losses.

We stress that the mechanism for the topological phase transition in these staggered helical lattices is distinctly different from the Floquet photonic topological insulator realized in the unstaggered helical lattice of  Ref.~\cite{supp_rechtsman2013}, which relied on the effective gauge field induced by the helical modulation. Indeed, the dimensionless effective gauge field strength $A_0 = a k_0 R_0 \Omega \sim 0.3$ is always small compared to the dimensionless quasienergy bandwidth $2\pi$. The reason why this small effective field is sufficient to induce a nontrivial phase in Ref.~\cite{supp_rechtsman2013} is that the unperturbed lattice already hosts Dirac points, so a nontrivial phase is induced by an infinitesimal perturbation. In contrast, phase transition displayed by the square helix lattice is intrinsically non-perturbative, akin to the topological transition at \emph{large} field strengths from $C=1$ to $C=-2$ predicted (but not observed) in Ref.~\cite{supp_rechtsman2013}.

The only way to close the quasienergy gap at $\phi = \pm \pi$ and reach the anomalous Floquet topological insulator phase using this small effective gauge field is with a modulation frequency $\Omega$ near-resonant with the unperturbed bandwidth $\sim t$, where $t \le 2.7$cm$^{-1}$ is the nearest neighbor coupling strength, set by $a$. In this resonant coupling regime, perturbation theory yields a gap size $\sim A_0$. The only way to tune the gap size is using the sole remaining free parameter, the helix radius $R_0$, which increases the bending loss exponentially~\cite{supp_bending_loss}. For example, the maximum practical value of $A_0 \sim 2.2$ from Ref.~\cite{supp_titum2015} requires $R_0 = 16\mu$m leading to prohibitive bending losses of 3dB/cm. For comparison, in the main text we achieved a topological gap size $\sim \pi$ with negligible bending losses and conservative waveguide parameters. 

The main difference between the staggered and unstaggered helical lattices is in the form of the modulation potential. The modulation in the latter is equivalent to a spatially uniform effective gauge field modulated in time at a single frequency $\Omega$. This harmonic driving conserves the Bloch momentum $\bk$ and can be easily understood in terms of local coupling in the frequency domain using a truncated repeated zone scheme~\cite{supp_rudner2013}. The phase transition is induced by coupling between neighboring zones in the frequency domain. In contrast, the step-like, on/off nature of the evanescent coupling terms in our staggered lattice is more naturally understood as a local modulation in the time domain: it involves many harmonics $n \Omega$ in the frequency domain, and the period-doubling induces scattering \emph{between} Bloch momenta $\bk$. Thus, a comparatively weak time modulation can be compensated by strong transverse scattering to induce a phase transition. 

\section{Floquet band structure calculation}

Here we explain our method for calculating the Floquet band structure of the continuum model. Waveguides fabricated using the femtosecond laser writing technique can be chosen to be single mode, with good confinement. Therefore, to a good approximation we can project the time-dependent field onto the bound waveguide modes, neglecting radiation losses (which are small, shown below self-consistently). This reduces the difficult problem of continuum Floquet band structure calculation to a low-dimensional discrete model, resembling the tight binding limit. 

\begin{figure}
\includegraphics[width=\columnwidth]{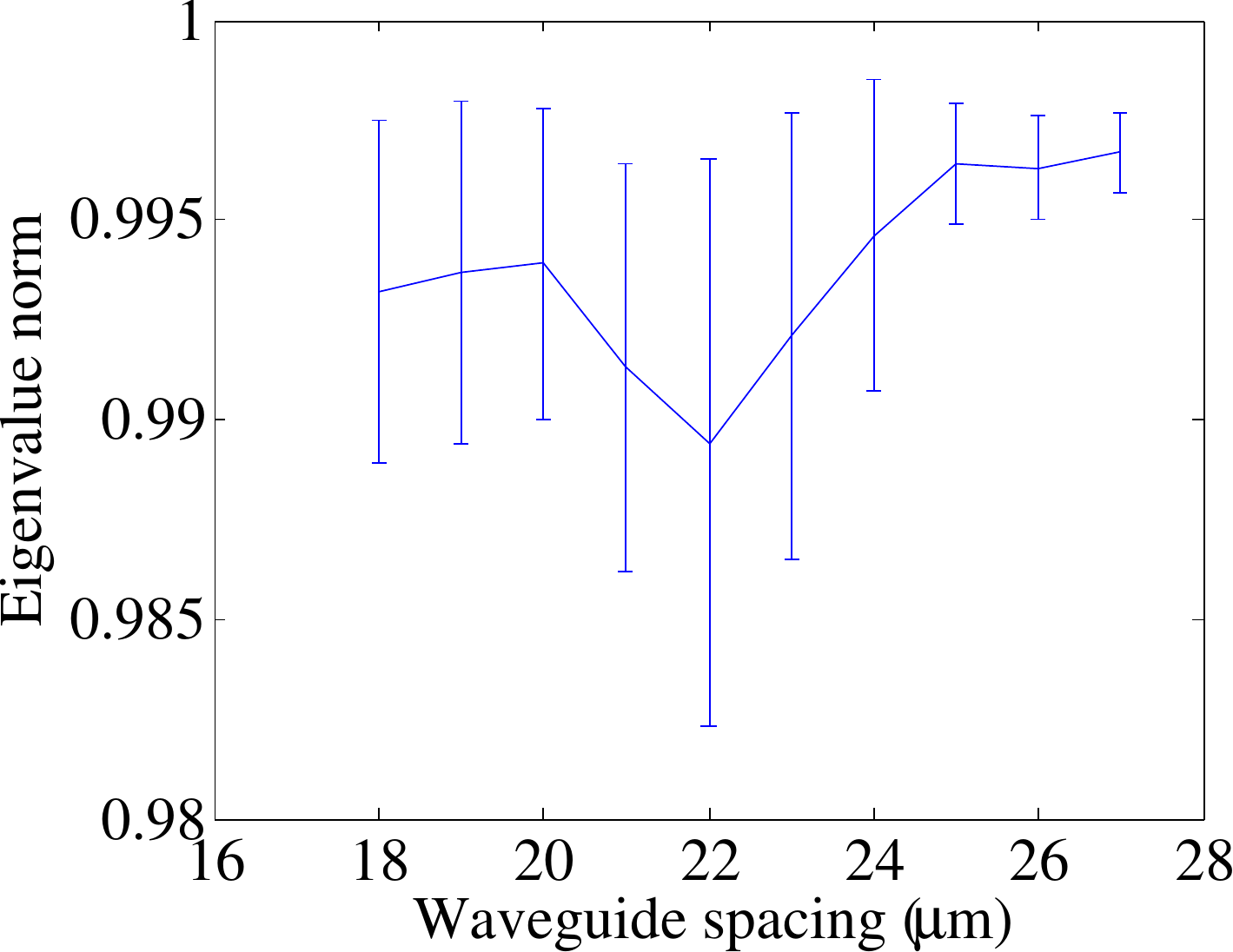}

\caption{Norm of the eigenvalues of the truncated evolution operator $\hat{U}(\bk)$ averaged over the Brillouin zone as a function of waveguide separation (symmetric lattice; $\alpha = 1$). Error bars denote one standard deviation over the ensemble of Bloch eigenmodes.}

\label{fig:losses}

\end{figure}

We obtain the bound modes by solving the \emph{static} Bloch mode eigenvalue problem at $z = 0$,
\be 
E_n (\bk ) \mid u_n (\bk ) \rangle = \hat{H}(\bk, 0) \mid u_n (\bk ) \rangle, 
\ee
where $\hat{H}(\bk, 0) = (\nabla + i \bk)^2 + V(\br, 0)$ is the Bloch Hamiltonian. For single mode waveguides, we can limit the number of bands $N$ to the number of waveguides per unit cell ($N=2$). This defines a truncated basis $\{ \mid u_n (\bk, 0) \rangle \}$, $n=1,...,N$. We numerically propagate each basis element for one modulation period, assuming twisted boundary conditions corresponding to the Bloch momentum $\bk$,
\begin{align}
\mid u_n (\bk, Z) \rangle &= e^{-i \int_0^{Z} \hat{H} (\bk,  z) dz} \mid u_n (\bk, 0 ) \rangle, \nonumber \\ &= \hat{U}(\bk, Z ) \mid u_n (\bk, 0 ) \rangle,
\end{align}
where $\hat{U}(\bk, Z )$ is the Bloch mode evolution operator. Since $\mid u_n (\bk, 0)\rangle$ is in general not an eigenstate of $\hat{H}(\bk, z\ne 0 )$, this evolution generically mixes the basis elements. We obtain a truncated evolution operator by projecting the final state back onto the basis states,
\be 
U_{mn} (\bk, Z ) = \langle u_m (\bk, 0 ) \mid \hat{U} (\bk, Z ) \mid u_n (\bk, 0 ) \rangle.
\ee
This method assumes that the evolution $\hat{U}$ does not couple modes beyond the $N$th band (including unbound radiative modes). If there is coupling to these modes, then the matrix $U_{mn}$ will be non-unitary. For our realistic choice of potential parameters, this quasistatic approximation  is good and we calculate the mean norm of the eigenvalues of $U_{mn}$ to exceed 0.99 in all cases, see Fig.~\ref{fig:losses}. This corresponds to an amplitude decay rate of $\sim 0.02$dB/cm. As $R_0 / Z \rightarrow 0$ the quasistatic approximation gets better and the norms approach unity.

Given $\hat{U}(\bk)$, we can compute its eigenvectors and hence evaluate the Chern number numerically~\cite{supp_chern_numerical}, and we can also determine the effective Hamiltonian via $\hat{H}_{\mathrm{eff}} = \frac{i}{\tau} \log (\hat{U} )$. Neglecting (non-Hermitian) loss terms, in the symmetric ($\alpha = 0$) lattice $\hat{H}_{\mathrm{eff}}$ is purely real at the high symmetry points $\bk = (0,0),(\pi,0),(0,\pi),(\pi,\pi)$, consistent with our symmetry argument above that the modulation-induced effective gauge field does not introduce a phase to the waveguide coupling - it merely renormalizes the coupling strength.

To fit the tight binding band structure to the continuum model's in Fig.~2 in the main text, we set $\Delta$ and $\theta_c$ to reproduce the band gap at $k_y = \pi/a$ and $k_y = 0$ respectively. This is already sufficient to demonstrate a reasonable agreement between the two models, although it obvious from Fig.~2(b) in the main text that for short lattice periods the functional form of the dispersion becomes qualitatively different (band edges are displaced from the $\Gamma$ point), which we attribute to corrections beyond the tight binding approximation (namely the coupling no longer occurs purely sequentially).

There is an intuitive and efficient way to obtain a correspondence between the continuum and tight binding model parameters: one simply has to simulate in the continuum model a single two-waveguide coupler to obtain its scattering matrix $\hat{S}_0$. $\Delta$ and $\theta_c$ can then be extracted using Eq.~\eqref{eq:s0}. Therefore the optimization of the design parameters for the anomalous Floquet insulator (eg. waveguide depths, separation, size) merely requires the optimization of a two-waveguide coupler, a relatively simple task. For example, for a symmetric coupler ($\Delta = 0$) with a given coupling constant $C$, one can obtain the required coupling angle $\theta_c$ by choosing an interaction length $L = \theta_c / C$.

\section{Conical diffraction}

Here we provide further information on the conical diffraction simulations of Fig.~4 in the main text, and discuss how they compare with the behavior of Dirac modes in regular honeycomb lattices.

\begin{figure}

\includegraphics[width=\columnwidth]{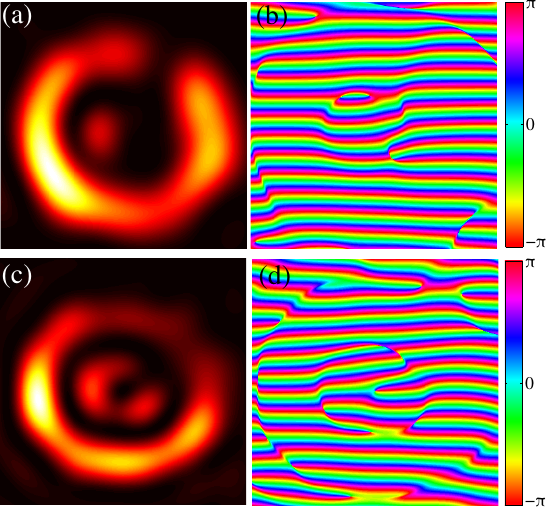}

\caption{Fourier filtering of conical diffraction. Raw intensity (a,c) and phase (b,d) profiles of conical (a,b) and discrete (c,d) conical diffraction. Global phase tilt of $e^{i \bs{K} \cdot \br}$ is extracted to obtain the smooth profiles shown in Fig.~4 in the main text.}

\label{fig:filtered_output}

\end{figure}

According to the effective Dirac Hamiltonian Eq.~\eqref{eq:dirac_hamiltonian}, the evolution of an initial ``pseudospin up'' state $\psi(\bk, 0) = g(\bk) (1,1)$ generates the final state
\begin{align}
\psi(L) &= g(\bk) e^{-i L \hat{H}_D} \psi(0), \\
&= g(\bk) \left[ \cos (k L) (1,1) - i \sin (kL ) e^{i \varphi} (1,-1) \right],
\end{align}
where $\bk = (k \cos \varphi, k \sin \varphi )$ is the momentum in polar coordinates, and $g(\bk)$ is the beam envelope. The ``pseudospin down'' (1,-1) component has a single charge phase vortex $e^{i \varphi}$, with chirality or handedness inherited from the Dirac Hamiltonian. To measure this vortex charge, we adapt the method from Ref.~\cite{supp_vortex_generation} by incorporating our sublattice degree of freedom into an expanded Brillouin zone. Namely, the ``down'' state exciting the two sublattices $\pi$ out of phase excites the four (equivalent) $\Gamma^{\prime}$ points at $\bk = \frac{2\pi}{a} (\pm 1, 0)$ and $\frac{2\pi}{a} (0,\pm 1)$. In experiment, one of these points can readily be isolated by Fourier filtering  the corresponding Bragg angle, say $\bs{K} = \frac{2\pi}{a}(0,1)$. We numerically implement this by applying a step filter to the output wavefunction $\psi_F = H ( \frac{\pi}{2a} - |\bk - \bs{K} | ) \psi ( L )$, where $H(\theta)$ is the Heaviside step function and we use a filter radius of $\frac{\pi}{2a}$ (the precise value is not important, as long as it is smaller than the Brillouin zone size $\frac{\pi}{a}$. The filtered output beams are plotted in Fig.~\ref{fig:filtered_output}. The intensity modulation induced by the lattice potential is smoothed out into smooth rings, and in the phase we observe a global tilt $e^{i \bs{K} \cdot \br}$ and dislocations corresponding to vortices. Subtracting the global tilt gives the smooth phase profiles plotted in the main text. In experiment an alternative way to measure the charge is via interference with a reference plane wave~\cite{supp_vortex_generation}. 

\begin{figure}

\includegraphics[width=\columnwidth]{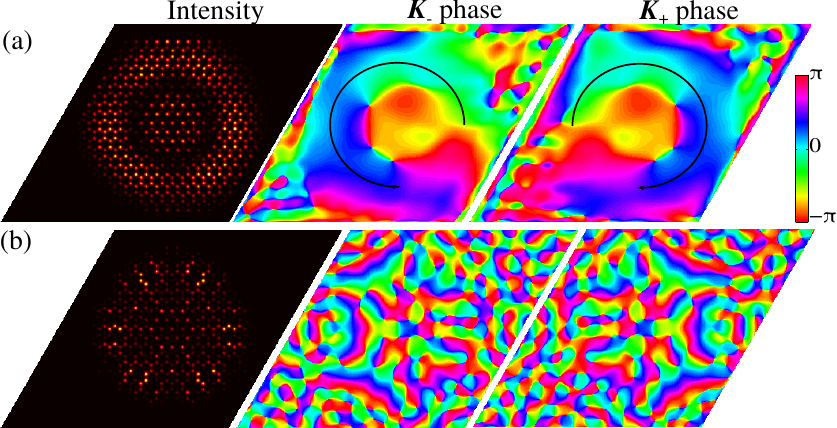}

\caption{Conical and discrete diffraction in the honeycomb lattice. (a) Output conical diffraction pattern when ``A'' sublattice is excited by a broad Gaussian beam. The charge of the filtered output phase profiles depends on which Dirac point $\bs{K}_{\pm}$ is excited. (b) Corresponding discrete diffraction pattern formed by excitation of a single ``A'' sublattice waveguide, and complex phase profiles obtained by filtering at each Dirac point.}

\label{fig:honeycomb_diffraction}

\end{figure}

We compare our results against conical and discrete diffraction in the honeycomb lattice in Fig.~\ref{fig:honeycomb_diffraction}. We use an initial state that excites only the ``A'' sublattice, which generates a pseudospin eigenstate. When this eigenstate is multiplied by a broad Gaussian envelope, the intensity of the resulting conical diffraction pattern shown in Fig.~\ref{fig:honeycomb_diffraction}(a) is insensitive to which of the Dirac points $\bs{K}_{\pm} = (\pm \frac{4 \pi}{3a},0)$ is selected. On the other hand, the filtered phase profiles do depend on Dirac point, displaying net charges of $\pm 1$. Thus in the honeycomb lattice, a pseudospin eigenstate is not intrinsically chiral - its chirality depends on which Dirac point is selected. Turning to discrete diffraction in Fig.~\ref{fig:honeycomb_diffraction}(b), the excitation of zero group velocity modes at the band edges leads to a more uniform output intensity distribution with no dark central spot, and the filtered phase profiles at the two Dirac points are sensitive to the filter width, and contain many vortex-antivortex pairs with no obvious chirality. 

\section{Disorder}

We introduce on-site disorder within the approximation of vanishing interaction length via a random detuning of the waveguide depths, replacing $\hat{S}_0$ in Eq.~\eqref{eq:s0} with
\begin{align}
\hat{S}_0 &= \left( \begin{array}{cc} e^{i (\Delta +V_{a,\bs{n}} )} & 0 \\ 0 & e^{-i (\Delta - V_{b,\bs{n}} )} \end{array} \right) \hat{U}_{\mathrm{coupler}},\nonumber \\
&=  \left( \begin{array}{cc} e^{i (\Delta +V_{a,\bs{n}} )} \cos \theta_c & -i e^{i (\Delta +V_{a,\bs{n}} )} \sin \theta_c  \\ -i e^{-i (\Delta - V_{b,\bs{n}} )} \sin \theta_c & e^{-i (\Delta - V_{b,\bs{n}} )}\cos \theta_c \end{array} \right),
\end{align}
where $V_{a,\bs{n}}$ and $V_{b,\bs{n}}$ are uniformly distributed random variables in the range $[-W/2,W/2]$. The disorder breaks translational symmetry, so we implement the model numerically using a finite lattice in real space with periodic boundary conditions.

In the weak antilocalization case shown in Fig.~5(a) of the main text, we take $W = \pi/80$.  In the strong-disorder case shown in Fig.~5(b) of the main text, we take $W = \pi/4$.  Note that the Anderson model of localization is only applicable to evolution governed by an effective Hamiltonian, the disorder is limited to $W \lesssim \pi / 2$, so that the phase accumulated over a modulation cycle does not exceed the quasienergy bandwidth.

To obtain the eigenmode participation number as a function of energy, we consider a lattice of size $N=30\times30$ unit cells and 20 realizations of the disorder, yielding an ensemble of 36,000 eigenmodes. We exactly diagonalize the evolution operator $\hat{U}$ to obtain its normalized eigenmode profiles $\psi_n$, where $n$ is the waveguide number. The participation number $P$ is defined as $P = 1/\sum_n |\psi_n|^4$. We divide the quasienergy $\beta$ into 51 bins and compute the mean value of each bin $P(\beta)$, yielding the plot in Fig.~5(b) of the main text.

\section{Robustness of edge states}

In this final section we present additional continuum model simulations of the unidirectional edge state propagation and its robustness against defect-induced backscattering. Like the Chern insulator, the anomalous Floquet insulator is a topological phase without any symmetries (the chiral modulation breaks the ``time reversal'' $T$ symmetry), so its edge modes should be robust against any kind of disorder, if one neglects effects such as absorption or bending losses into continuum modes.

\begin{figure*}

\includegraphics[width=2\columnwidth]{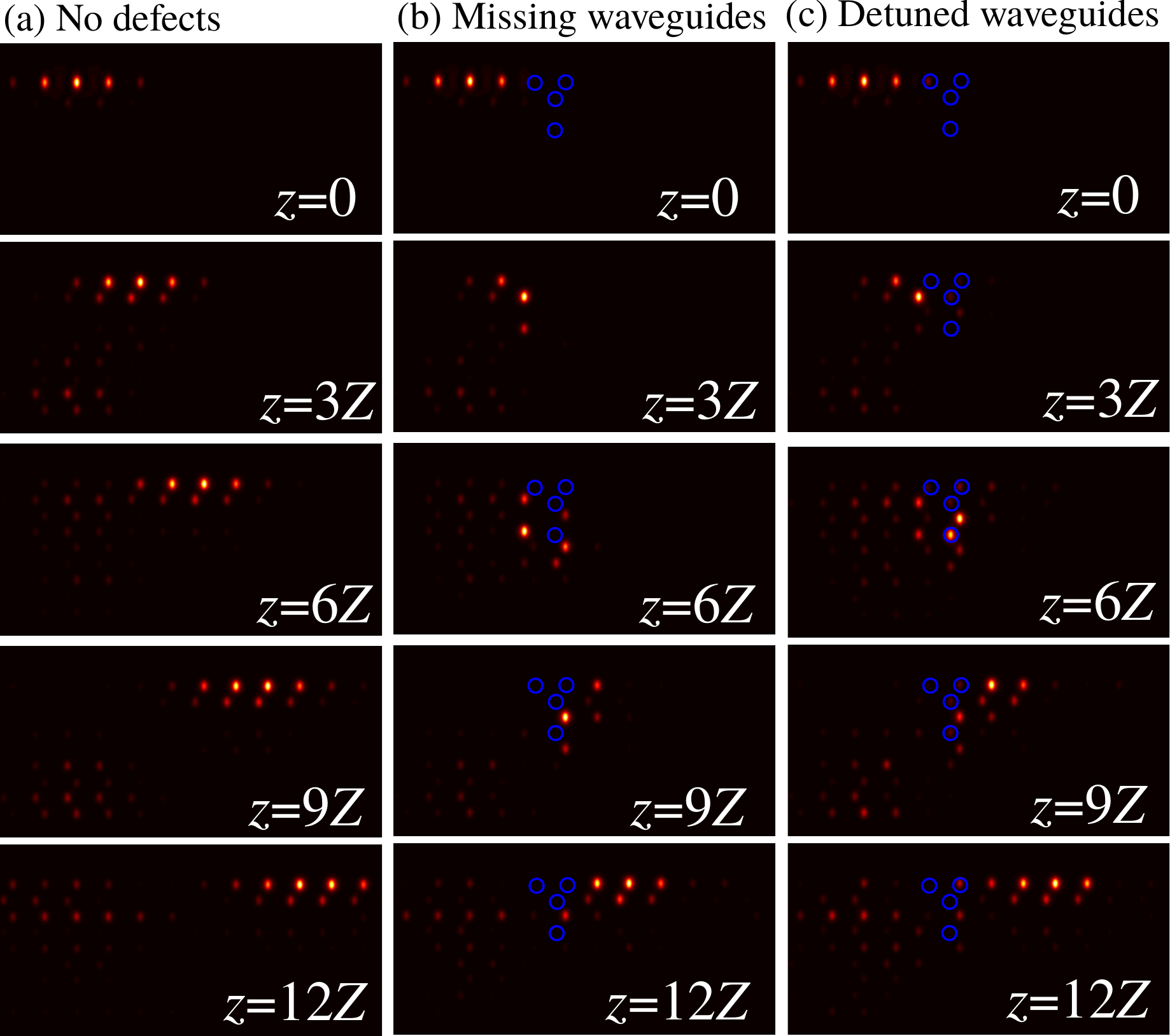}

\caption{Robustness of unidirectional edge propagation against scattering by a defect. Lattice period $a=20\mu$m (nontrivial phase), other lattice parameters same as in main text. Panels show intensity profiles after propagating a distance $z$. (a) Semi-infinite strip without defects. (b) Waveguides at circled positions removed. (c) Waveguides at circled positions detuned.}

\label{fig:defect_scattering}

\end{figure*}

We consider a semi-infinite strip in the nontrivial phase, using the same lattice parameters as in Fig. 2(b) of the main text. Fig.~\ref{fig:defect_scattering} shows intensity profiles after various propagation distances $z$ when a few waveguides at the edge are excited. In the defect-free lattice in Fig.~\ref{fig:defect_scattering}(a) most of the light remains confined to the edge, propagating to the right. Some bulk modes are also weakly excited and are visible as a small fraction of the beam spreading into the bulk. When a defect is introduced by removing some of the edge waveguides in Fig.~\ref{fig:defect_scattering}(b) the wavepacket travels around the defect without backscattering, although it is delayed with respect to the defect-free case. Similar backscattering-free propagation is observed in Fig.~\ref{fig:defect_scattering}(c), where the depths of the indicated waveguides are detuned up to 20\%.

\end{document}